\documentclass[review]{elsarticle}

\usepackage{lineno,hyperref}
\modulolinenumbers[5]

\usepackage{filecontents}
\usepackage{csvsimple}
\usepackage{tikz}
\usepackage{pgfplots}
\usepackage{mathtools}
\usepackage{placeins}
\usepackage{multirow}
\usepackage{float}
\usepackage[utf8]{inputenc}
\usepackage{xcolor}
\usepackage{makecell}
\usepackage{rotating}
\usepackage{graphicx}
\usepackage{threeparttable}
\usepackage{wasysym}
\usepackage{algorithm}
\usepackage{algpseudocode}
\usepackage{afterpage}
\usepackage{array}
\usepackage{booktabs}
\usepackage{lscape}
\usepackage{gensymb}

\usepackage[labelformat=simple]{subcaption}

\usepgfplotslibrary{groupplots}
\pgfplotsset{compat=1.13}

\tikzset{trim axis left,trim axis right}

\newcolumntype{L}[1]{>{\raggedright\let\newline\\\arraybackslash\hspace{0pt}}m{#1}}
\newcolumntype{C}[1]{>{\centering\let\newline\\\arraybackslash\hspace{0pt}}m{#1}}
\newcolumntype{R}[1]{>{\raggedleft\let\newline\\\arraybackslash\hspace{0pt}}m{#1}}

\newcommand*\rot{\rotatebox{90}}

\journal{arxiv.org}

\biboptions{authoryear}

\begin{document}

\begin{frontmatter}

	\title{A rasterized ray-tracer pipeline for real-time, multi-device sonar simulation}

	%% Authors:
	\author[senai,ufba]{Rômulo Cerqueira\corref{cor1}} 	\ead{romulo.cerqueira@ufba.br}
	\author[senai]{Tiago Trocoli} 					\ead{trocolit@gmail.com}
	\author[kraken]{Jan Albiez}							\ead{jan@ankerwin.de}
	\author[ufba]{Luciano Oliveira}					\ead{lrebouca@ufba.br}
	\cortext[cor1]{Corresponding author: Rômulo Cerqueira.}
    \ead{romulo.cerqueira@ufba.br}

	%% Filiations:
	\address[senai]{Brazilian Institute of Robotics, SENAI CIMATEC, Salvador, Bahia, Brazil}
	\address[kraken]{Kraken Robotik GmbH, Bremen, Germany}
	\address[ufba]{Intelligent Vision Research Lab, Federal University of Bahia, Salvador, Bahia, Brazil}

	\begin{abstract}
		Simulating sonar devices requires modeling complex underwater acoustics, simultaneously rendering time-efficient data. Existing methods focus on basic implementation of one sonar type, where most of sound properties are disregarded. In this context, this work presents a multi-device sonar simulator capable of processing an underwater scene by a hybrid pipeline on GPU: Rasterization computes the primary intersections, while only the reflective areas are ray-traced. Our proposed system launches few rays when compared to a full ray-tracing based method, achieving a significant performance gain without quality loss in the final rendering. Resulting reflections are then characterized as two sonar parameters: Echo intensity and pulse distance. Underwater acoustic features, such as speckle noise, transmission loss, reverberation and material properties of observable objects are also computed in the final generated acoustic image. Visual and numerical performance assessments demonstrated the effectiveness of the proposed simulator to render underwater scenes in comparison to real-world sonar devices.
	\end{abstract}


\end{frontmatter}

% \linenumbers

% =======================================================================================

\section{Introduction}
\label{intro}

The number of underwater structures in the offshore industry has significantly increased over the last decades, and so the need of monitoring, inspection and intervention of these structures \citep{wang2018}. Since autonomy is necessary to reduce mission expenses, the offshore industry has leading the development of autonomous underwater vehicles (AUVs) to accomplish the main field tasks. With a pre-programmed mission and onboard sensors, AUVs are able to perform completely autonomous decisions, returning to surface only for servicing.

The accomplishment of AUV tasks demands to deal with challenges inherent to undersea environment. For instance, beneath the water, optical cameras are affected by turbidity and lightning conditions, thus restricting the image quality to short visible ranges. On the other hand, imaging sonars take advantage of the low attenuation of sound waves in order to cover larger areas than those ones covered by optical cameras, although producing noisy data with low resolution.

AUV real-world experimentation is challenging, mainly due to human resources, time consumption and hazards involved on deployment and testing the underwater vehicles in the target domain. While initial experiments can be performed in water tanks (\textit{e.g.}, low-level control and basic prototyping), high-level tests require trials in deep open waters (\textit{e.g.}, way-point navigation, mapping and autonomous control). An unexpected behavior of an AUV may result in an unrecoverable equipment, causing a considerable financial loss. This way, simulation of underwater sensors and reproducible environments is essential to cope with insufficient data, as well as to develop effective algorithms before tests in the wild.

To contribute with the development of underwater acoustic-based systems, this paper introduces a novel simulator able to reproduce the operation of different sonar devices. Rendering of a virtual scene is accelerated by a selective rasterization and ray-tracing scheme on GPU, where the computational resources are allocated only for reflective regions. Subsequently, the resulting reflections are converted to the acoustic scene representation on CPU, including several phenomena present on the sonar images.

% =======================================================================================

\subsection{Related work}
\label{intro:related}

% Introduction
By considering the complexity in the process of transmitting sound through the water, several mathematical and computational models have been proposed to approximate the calculation of acoustic propagation \citep{etter2018}. Ray-based methods are the most common solutions to simulate underwater sonar systems \citep{bell1997,gueriot2007,gu2013,kwak2015,demarco2015,sac2015,mai2018,soares2016}, although other approaches can also be considered \citep{coiras2009,cerqueira2017,gwon2017}. All simulation methods try to mimic one or more types of sonar devices.

% Side-scan sonar
\textbf{Side scan sonar (SSS) simulation}: \cite{bell1997} presented a simulator for SSS imagery based on optical ray-tracing, where a group of rays is projected to insonify the scene and produce the acoustic data; fractal models are used to represent the roughness surface of the seafloor; stochastic influences as noise and reverberation are neglected in that work. Instead of propagating many individual rays, \cite{gueriot2007} developed a volume-based approach with a tube tracing technique; the tubes are composed of four rays, which intersect a certain area to allow computing the backscattered energy; the few launched rays optimized the sonar rendering, while the surface details and transmitting signal characteristics are suppressed. By using a frequency domain-based method, \cite{coiras2009} produced frames from a virtual SSS by using Fourier transform; the returned intensity relies on the angle of incidence applied to a basic Lambert illumination model; physical effects, such as noise and multi-path returns, are considered, although the method was not designed to operate online. With a simplified Lambert diffusion model, \cite{gwon2017} generated SSS data integrated with UWSim simulator and ROS framework; acoustic frames are degraded with speckle (low frequency) and Rayleigh (high frequency) noises; although the performance of feature matching methods decays in images containing multiplicative noise, due to the variance for intensity and affine changes, the authors applied SIFT, SURF, ORB and AKAZE algorithms to evaluate the similarity between two consecutive frames, obtaining a very low number of inliers for all feature extractors.

% Forward-looking sonar
\textbf{Forward-looking sonar (FLS) simulation}: \cite{gu2013} modeled an FLS system, where the rays are comprised of basic lines, equivalent to the number of pixels of sonar image to be emulated; the reflection representation is severally reduced to three colors only (black, gray and white). \cite{kwak2015} improved Gu \textit{et al.}'s method by introducing sound attenuation effect in order to produce gray-scale sonar images; by assuming a mirror-like reflection model, the sonar system only considers specular reflections, so that the method is only successful for smooth surfaces. \cite{sac2015} introduced an acoustic model by combining ray-tracing in frequency domain; the intensity and the range of sonar data are calculated by Lambert diffusion model and Euclidean distance respectively; the high average time to compute a single FLS frame prevents the use of the method in real-time operations. \cite{demarco2015} detailed an FLS simulator integrated with Gazebo simulator and ROS for diving assistance; the ray path mimics the sound wave to generate a point cloud; the simulated images are compared with real ones, although the reflectivity of the objects and the noise models are analytically defined. \cite{mai2018} conceived a simulator based on ray propagation to produce acoustic data; by assuming only the freshwater component, the sound attenuation is partially considered, while other physical properties of sound are ignored; time consumption to calculate one single frame has not been well established by \cite{mai2018}.

% Mechanical scanning imaging sonar
\textbf{Mechanical scanning imaging sonar (MSIS) simulation}: \cite{soares2016} fused the ray-tracing and additive noise models, proposed in \cite{bell1997} and \cite{coiras2009}, respectively, to produce single beam data; in that work, no image distortion induced by robot movement was considered; the simulated frames were later used to feed an underwater localization system based on Hilbert maps. \cite{cerqueira2017} introduced a GPU-based simulator to reproduce the operation of two sonar devices; by deferred shading, the rasterization rendering was exploited to compute the acoustic parameters (\textit{i.e.}, echo intensity, pulse distance and azimuth angle); sound phenomena such as multiplicative noise and material properties were addressed, while multipath returns, attenuation and additive noise did not; experiments comparing real-world acoustic images certified the use of the simulator by real-time applications.

\begin{table*}[!t]
	\centering
	\scriptsize
	\addtolength{\leftskip} {-1cm}
	\addtolength{\rightskip}{-1cm}
	\newcommand*\f[1]{\ifcase#1 \Circle\or\LEFTcircle\or\CIRCLE\fi}
	\caption{Summary of state-of-the-art works on imaging sonar simulation.}
	\begin{center}
	\begin{tabular}{|c|l|c|c|c|c|c|c|c|c|c|c|c|c|}
		\cline{3-14}
		\multicolumn{2}{c|}{} & \multicolumn{12}{c|}{\textbf{Works~}} \\
		\cline{3-14}
		\multicolumn{2}{c|}{}
		& \rot{\cite{bell1997}}
		& \rot{\cite{gueriot2007}}
		& \rot{\cite{coiras2009}}
		& \rot{\cite{gu2013}}
		& \rot{\cite{kwak2015}}
		& \rot{\cite{sac2015}}
		& \rot{\cite{demarco2015}}
		& \rot{\cite{soares2016}}
		& \rot{\cite{gwon2017}}
		& \rot{\cite{cerqueira2017}}
		& \rot{\cite{mai2018}}
		& \rot{Ours} \\
		\hline
		\multirow{3}{*}{\rot{\textbf{Type~}}}
			& SSS & \f2 & \f2 & \f2 & \f0 & \f0 & \f0 & \f0 & \f0 & \f2 & \f0 & \f0 & \f0 \\
			\cline{2-14}
			& FLS & \f0 & \f0 & \f0 & \f2 & \f2 & \f2 & \f2 & \f0 & \f0 & \f2 & \f2 & \f2 \\
			\cline{2-14}
			& MSIS & \f0 & \f0 & \f0 & \f0 & \f0 & \f0 & \f0 & \f2 & \f0 & \f2 & \f0 & \f2 \\
			\hline
		\multirow{4}{*}{\rot{\textbf{Model~}}}
			& Frequency domain &  \f0 & \f0 & \f2 & \f0 & \f0 & \f2 & \f0 & \f0 & \f0 & \f0 & \f0 & \f0 \\
			\cline{2-14}
			& Tube tracing & \f0 & \f2 & \f0 & \f0 & \f0 & \f0 & \f0 & \f0 & \f0 & \f0 & \f0 & \f0 \\
			\cline{2-14}
			& Ray-tracing & \f2 & \f0 & \f0 & \f2 & \f2 & \f2 & \f2 & \f2 & \f0 & \f0& \f2 & \f2 \\
			\cline{2-14}
			& Rasterization &  \f0 & \f0 & \f0 & \f0 & \f0 & \f0 & \f0 & \f0 & \f0 & \f2 & \f0 & \f2 \\
		\hline
		\multirow{7}{*}{\rot{\textbf{Features~}}}
			& Reflection model & \f2 & \f2 & \f2 & \f0 & \f1 & \f2 & \f1 & \f2 & \f1 & \f2 & \f2 & \f2 \\
			\cline{2-14}
			& Surface irregularities & \f2 & \f0 & \f0 & \f0 & \f0 & \f0 & \f0 & \f0 & \f0 & \f2 & \f0 & \f2 \\
			\cline{2-14}
			& Surface reflectance & \f0 & \f0 & \f0 & \f0 & \f0 & \f0 & \f1 & \f0 & \f0 & \f2 & \f0 & \f2 \\
			\cline{2-14}
			& Attenuation & \f2 & \f0 & \f0 & \f0 & \f1 & \f0 & \f0 & \f0 & \f0 & \f0 & \f1 & \f2 \\
			\cline{2-14}
			& Speckle noise & \f0 & \f0 & \f1 & \f0 & \f0 & \f0 & \f1 & \f1 & \f1 & \f1 & \f0 & \f2 \\
			\cline{2-14}
			& Reverberation & \f0 & \f0 & \f1 & \f0 & \f0 & \f1 & \f0 & \f1 & \f0 & \f0 & \f0 & \f1 \\
			\cline{2-14}
			& Robotics framework support & \f0 & \f0 & \f0 & \f0 & \f0 & \f0 & \f2 & \f0 & \f0 & \f2 & \f2 & \f2 \\
			\hline
		\multirow{3}{*}{\rot{\textbf{Eval.~}}}
			& Qualitative & \f2 & \f2 & \f2 & \f2 & \f2 & \f2 & \f2 & \f2 & \f2 & \f2 & \f2 & \f2 \\
			\cline{2-14}
			& Computation time & \f0 & \f0 & \f0 & \f0 & \f0 & \f1 & \f1 & \f0 & \f0 & \f2 & \f1 & \f2 \\
			\cline{2-14}
			& Quantitative & \f0 & \f0 & \f0 & \f0 & \f0 & \f0 & \f0 & \f0 & \f0 & \f2 & \f0 & \f2 \\
		\hline
	\end{tabular}
	\begin{tablenotes}
		\centering
		\item \f2 = \text{provides property}; \f1 = \text{partially provides property}; \f0 = \text{does not provide property}.
	\end{tablenotes}
	\label{tab:comparison}
	\end{center}
\end{table*}

% =======================================================================================

\subsection{Contributions}
\label{intro:contrib}

This paper proposes a sonar simulator that extends the work in \citep{cerqueira2017} by combining rasterization and ray-tracing to optimize acoustic reflections and fulfilling the missing physical phenomena. A comparative summary between the state-of-the-art works and ours is detailed in Table \ref{tab:comparison}.

Instead of simulating a specific sonar type \citep{bell1997,gueriot2007,coiras2009,gu2013,demarco2015,kwak2015,sac2015,soares2016,gwon2017,mai2018}, our proposed method is able to reproduce the operation of FLS and MSIS sensors. A selective rasterized ray-tracer is integrated on GPU, where the computational resources are restricted to only reflective regions; this combination enables multipath reflections (not present in rasterization-based works, such as in \cite{cerqueira2017}), launching few rays with the same final result in comparison with full ray-tracing and tube tracing methods \citep{bell1997,gueriot2007,gu2013,demarco2015,kwak2015,sac2015,soares2016,gwon2017,gueriot2007,mai2018}. Additionally the number of intersection tests of ray-tracing model is significantly reduced by using bounding volumes and a ray-box intersection algorithm, accelerating the rendering time as a consequence.

The sonar simulator is already integrated with a robotics framework (\textit{i.e.}, Rock), supporting the integration with real and simulated robotic systems, feature present in \citep{demarco2015,cerqueira2017,mai2018}. The echo intensity from observable objects depends on surface normal directions, material reflectivity and sound attenuation properties, differently from existing approaches \citep{gu2013,demarco2015,gwon2017}, where the reflection value is empirically defined. Yet, the reflection model is valid for any type of surface representation, in opposition to \citep{demarco2015,sac2015}. Five of the analyzed works consider either additive or multiplicative noise, while speckle effect is just partially simulated \citep{coiras2009,sac2015,soares2016,gwon2017,cerqueira2017}. In our work, speckle noise is fully reproduced.

Our experiments comprises qualitative, computational time and quantitative evaluations between simulated and real-world sonar data, assessing time-efficiency and rendering quality of the generated acoustic images.

% =======================================================================================

\section{Working with underwater sonars}
\label{sonar:intro}

% Active and Passive sonars
Sonar systems use the propagation of sound waves to detect and locate objects underwater. These systems are grouped into two basic types: Passive and active \citep{rossing2015}. A passive sonar essentially listens for the sound waves made by submerged objects; in contrast, an active sonar transmits sound pulses, and then listens for echoes. Imaging sonars are classified as active devices.

\begin{figure}[t]
	\centering
	\includegraphics[width=1\columnwidth]{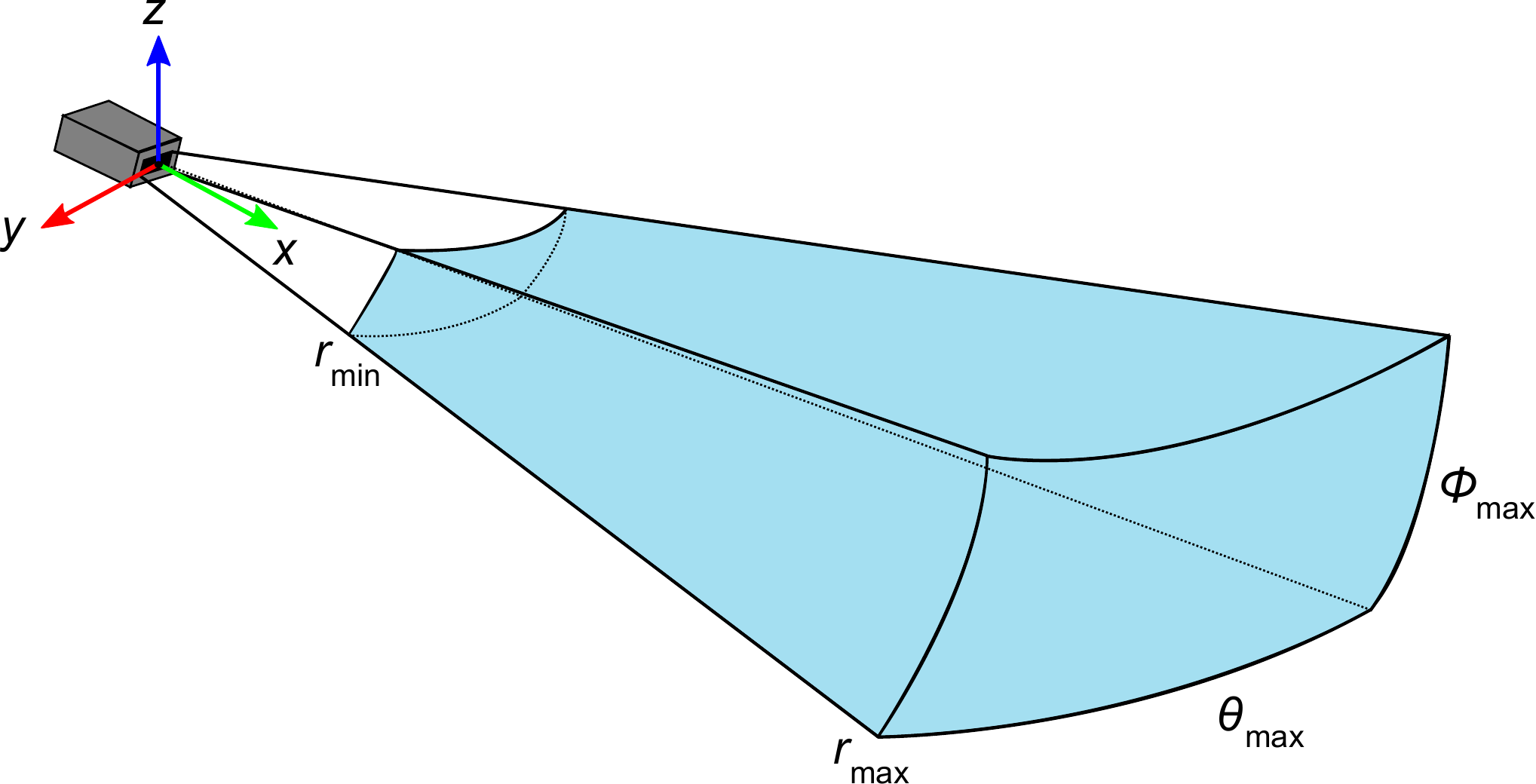}
    \caption{Viewing volume of imaging sonar. The observable scene is defined by the minimum and maximum ranges $r_{min}$ and $r_{max}$, maximum azimuth angle $\theta_{max}$, and maximum elevation angle $\phi_{max}$.}
    \label{fig:sonar_covered_area}
\end{figure}

% How active sonar works
To compose an acoustic image, an active sonar insonifies the scene with a sound wave. The visible area is delimited by maximum azimuth angle $\theta_{max}$, maximum elevation angle $\phi_{max}$, and minimum $r_{min}$ and maximum $r_{max}$ ranges, as illustrated in Fig. \ref{fig:sonar_covered_area}. In case that a sound wave hits an object, the returning echo is sampled as a function of range and bearing, since the speed of sound in water is known. The transducer reading in a given direction composes a \textit{beam}, while each distance sampled along this beam is named \textit{bin}. The strength of backscattered energy in each bin determines the echo intensity from an insonified object. Combining the array of transducer readings, the group of echo intensities forms an image of the reflective surfaces in front of the sonar head.

% =======================================================================================

\subsection{Geometry of the sonar projection}
\label{sonar:geometry}

\begin{figure}[t]
	\centering
    \includegraphics[width=1\columnwidth]{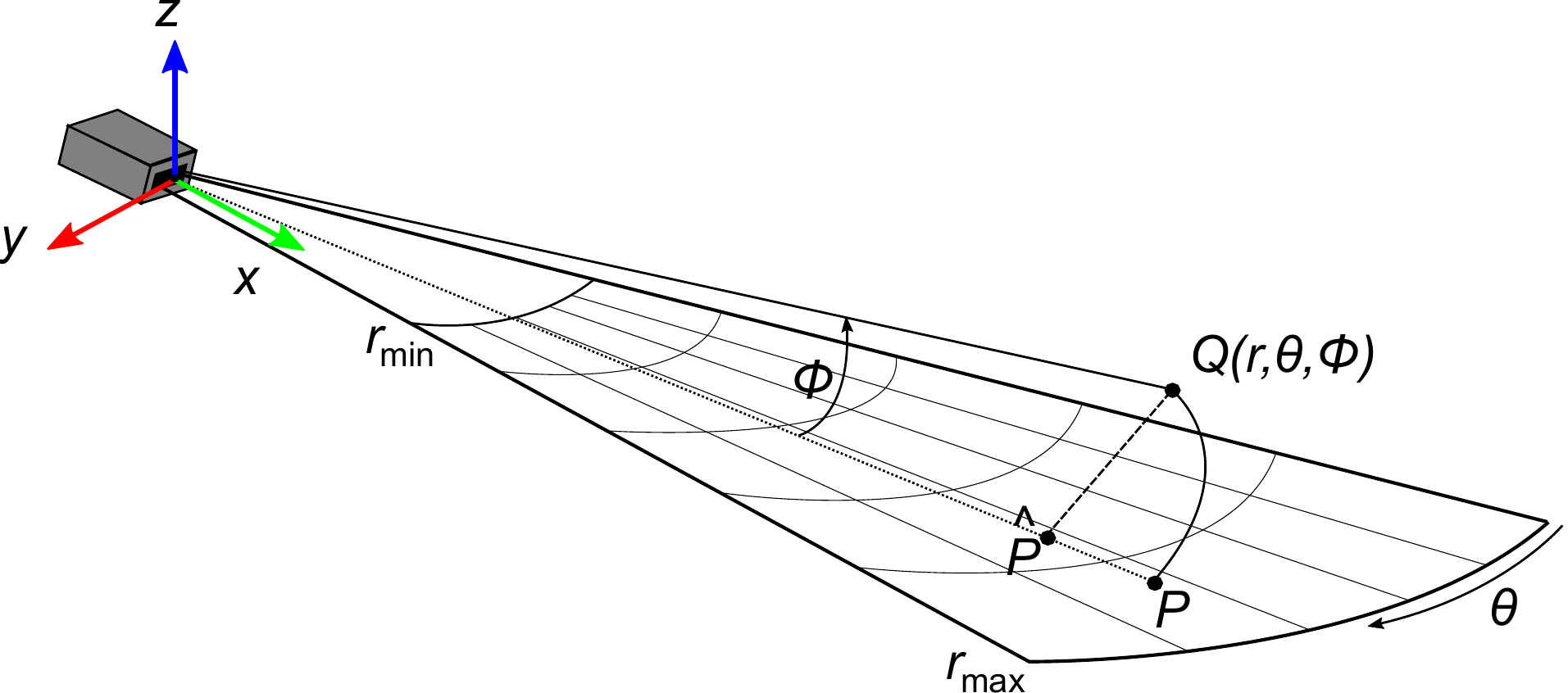}
	\caption{Model of the imaging sonar projection. A spherical point $Q(r,\theta,\phi)$ is projected into a point $P$ on an image plane. Considering an orthographic approximation, the point $P$ is mapped onto $\hat{P}$, which is equivalent to all points along the same arc.}
	\label{fig:sonar_projection}
\end{figure}

% Sonar projection model
A 3D point is usually expressed in Cartesian coordinates as $[x, y, z]^{T}$. A sonar system has the reference frame defined in spherical coordinates as $Q = [r, \theta, \phi]^{T}$, with range $r$, azimuth angle $\theta$, and elevation angle $\phi$, as depicted in Fig. \ref{fig:sonar_projection}. The conversion of Cartesian to spherical coordinates is given by

\begin{equation}
    \label{eq:geom:b}
    Q = \begin{bmatrix}
		r \\
		\theta \\
		\phi
	  \end{bmatrix} =
	  \begin{bmatrix}
		\sqrt{x^2 + y^2 + z^2} \\
		\tan^{-1}(y/x)\\
		\tan^{-1}(\sqrt{x^2 + y^2}/z)
	  \end{bmatrix} \, .
\end{equation}

Since the elevation angle $\phi$ is missing during the process of acoustic projection, the sonar system measures the range $r$ and azimuth angle $\theta$ onto the zero-elevation plane, as an approximation to an orthographic projection \citep{johannsson2010}. This two-dimensional system is named polar coordinates and follows a nonlinear model defined as

\begin{equation}
    \label{eq:geom:c}
    \hat{P} = \begin{bmatrix}
		x \\
		y
	  \end{bmatrix} =
	  \begin{bmatrix}
		r \cos{\theta} \\
		r \sin{\theta}
	  \end{bmatrix} \, .
\end{equation}

% =======================================================================================

\subsection{Sound attenuation}
\label{sonar:attenation}

When a sound pulse propagates through the water, the acoustic energy is gradually converted into heat by a spherical spreading, absorption and chemical properties of the sea. This effect decreases the signal amplitude exponentially with distance, and the total acoustic attenuation in the ocean is expressed by three additive components: Relaxation of borid acid ($H_{3}BO_{3}$) molecules below $1$ kHz, relaxation of magnesium sulphate ($MgSO_{4}$) below 100 kHz, and viscosity of pure water \citep{bjorno2017}. A common attenuation method is proposed by \cite{ainslie1998}, where the attenuation coefficient is expressed as

\begin{equation}
	\label{eq:atten:a}
	\alpha = \alpha_{B} + \alpha_{M} + \alpha_{F} \, ,
\end{equation}
with the boric acid component $\alpha_{B}$ defined as

\begin{equation}
	\label{eq:atten:b}
	\alpha_{B} = 0.106\frac{f_{1}f^{2}}{f^{2}+f_{1}^2}e^{(pH-8)/0.56} \, ,
\end{equation}

\begin{equation}
	\label{eq:atten:b2}
	f_{1} = 0.78\left(\frac{S}{35}\right)^{1/2}e^{T/26} \, ,
\end{equation}

the magnesium sulphate component $\alpha_{M}$ defined as

\begin{equation}
	\label{eq:atten:c}
	\alpha_{M} = 0.52\left(1+\frac{T}{43}\right)\left(\frac{S}{35}\right)\frac{f_{2}f^2}{f^2+f_{2}^2}e^{-z/6} \, ,
\end{equation}

\begin{equation}
	\label{eq:atten:c2}
	f_{2} = 42e^{T/17} \, ,
\end{equation}

and the freshwater component $\alpha_{F}$ defined as

\begin{equation}
	\label{eq:atten:d}
	\alpha_{F} = 0.00049f^2e^{-(T/27 + z/17)} \, ,
\end{equation}
where $\alpha$ is the intensity absorption coefficient in dB/km, $f$ is frequency in kHz, $S$ is salinity in parts per thousand (ppt), $pH$ is acidity, $z$ is depth in km, and $T$ is the water temperature in Celsius.

% =======================================================================================

\subsection{Speckle noise}
\label{sonar:noise}

Due to coherent nature of scattering phenomena, sonar images are affected by speckle noise, a granular pattern, which severely deteriorates the visual quality, and reduces relevant features as edges and shapes. This type of noise produces random variations of image intensity, which causes light and dark pixels and interferes in further operations, such as object detection and segmentation. The noisy image, $\hat{I}$, has been expressed as \citep{mateo2009}

\begin{equation}
	\label{eq:noise}
	\hat{I}(r,\theta) = I(r,\theta) \eta_{m}(r,\theta) + \eta_{a}(r,\theta) \, ,
\end{equation}
where $(r,\theta)$ are the polar coordinates, $I$ is the noise-free image, and $\eta_{m}$ and $\eta_{a}$ are the multiplicative and additive noise components, respectively.

% =======================================================================================

\subsection{Reverberation}
\label{sonar:reverb}

When active sonars transmit sound pulses, incoming echoes are usually returned from several different sources. The result of unwanted echoes is named reverberation, which is mainly caused by the multiple path propagation and successive interactions of the transmitted signal, weakening the sound intensity \citep{hodges2010}. Sources of reverberation in the ocean include the surface, the seafloor and the volume of water.

% =======================================================================================

\subsection{Representation of sonar image}
\label{sonar:image}

\begin{figure}[t]
	\centering
	\begin{subfigure}[b]{0.57\linewidth}
    \includegraphics[width=\textwidth,height=52mm]{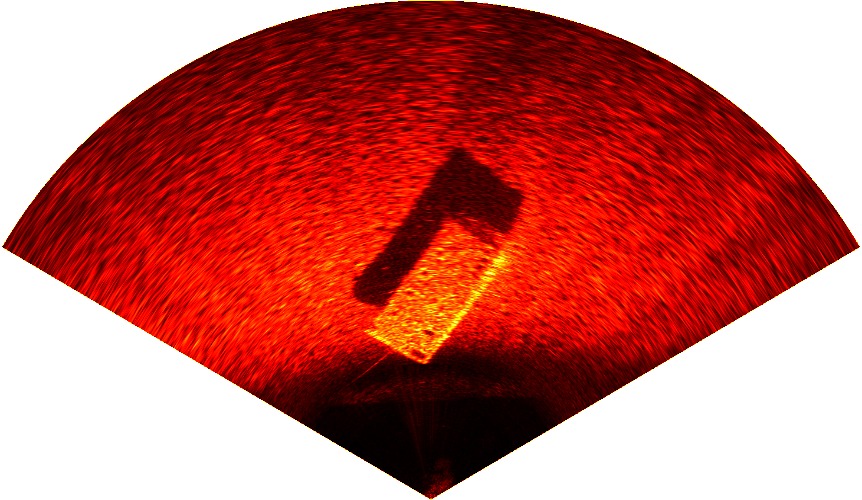}
    \caption{Raw data as Cartesian image.}
    \label{fig:view_cart}
	\end{subfigure}
	\begin{subfigure}[b]{0.42\linewidth}
		\centering
		\includegraphics[width=\textwidth,height=35mm,angle=90]{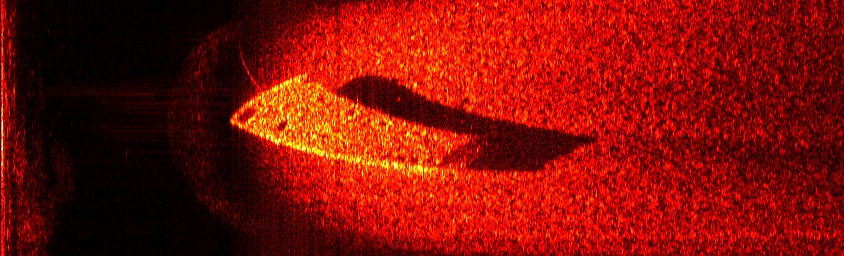}
    \caption{Raw data as polar image.}
    \label{fig:view_polar}
  \end{subfigure}

  \caption{Different types of acoustic data representations. A wrecked ferry was captured with FLS Tritech Gemini 720i sensor from a real AUV. In this paper, the simulated images are displayed in Cartesian coordinates to retain the characteristics of the insonified objects \subref{fig:view_cart}, while the polar image is applied during similarity evaluation without loss of original data \subref{fig:view_polar}.}
	\label{fig:view_representation}
\end{figure}

The raw sonar data is generated as a function of range and bearing, resulting in a polar image $I(r,\theta)$ with an echo intensity value for each pixel. For a better human interpretation, this polar representation can be converted to Cartesian coordinates $I(x,y)$ using Eq. (\ref{eq:geom:c}). In Cartesian coordinates, the fan-shaped image preserves the target geometry. The conversion from polar to Cartesian system entails a non-uniform resolution, where the representation of the closest bins to sonar origin are superscripted, while the far ones are interpolated, yielding to image distortions and object flatness. The raw polar data and the corresponding representation in Cartesian space is illustrated in Fig. \ref{fig:view_representation}.

% =======================================================================================

\section{Simulating acoustic images based on a rasterized ray-tracing pipeline}
\label{dev:intro}

\begin{sidewaysfigure}
	\centering
  \includegraphics[width=1\linewidth]{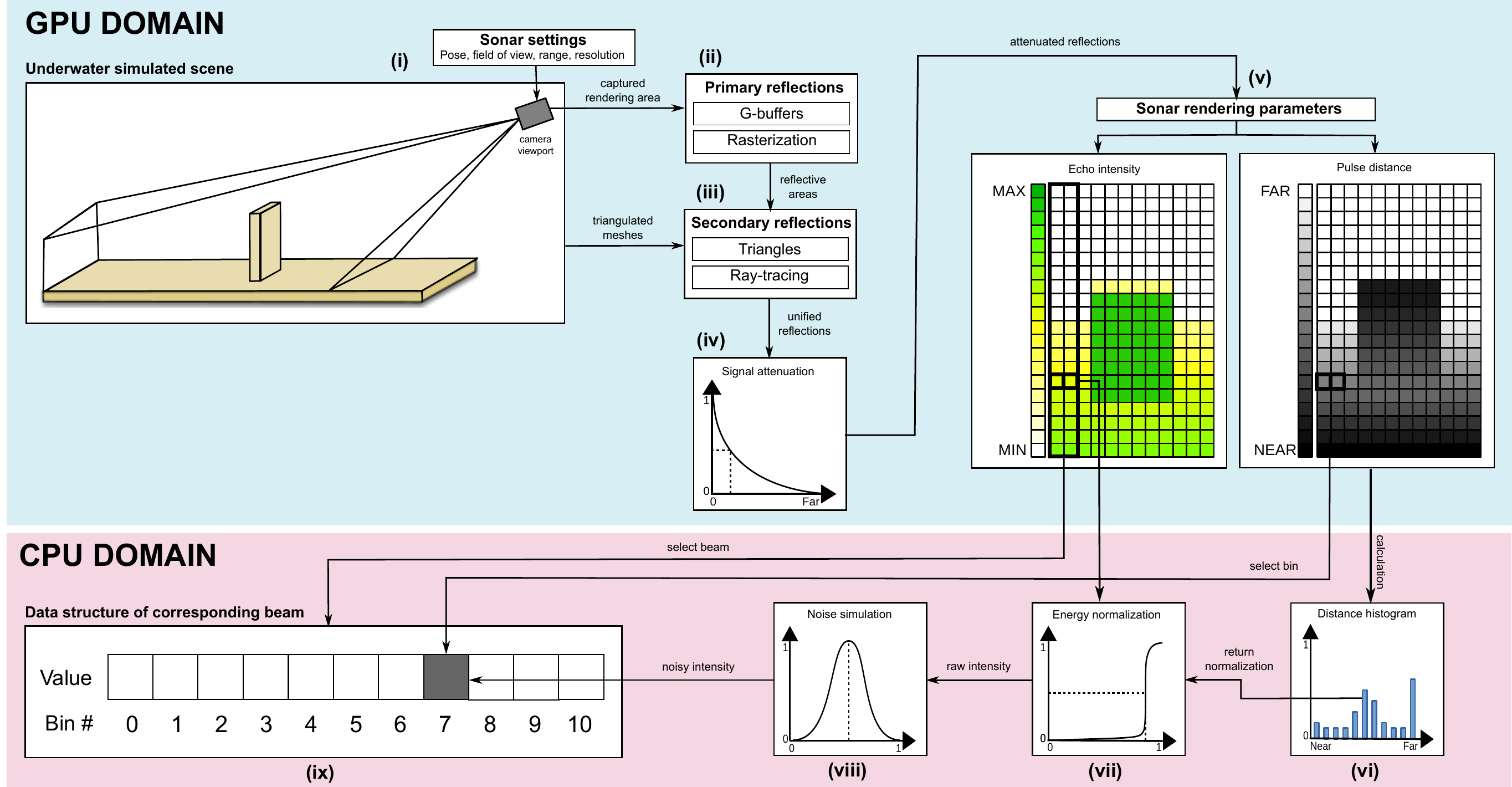}
	\caption{Overview of proposed imaging sonar simulation. On GPU domain: (i) a virtual camera captures the observable scene; (ii) by rasterization, the primary reflections are computed by using the surface normal and depth values from G-buffers; (iii) only the reflective areas are ray-traced for secondary reflections; (iv) the signal attenuation model decays the amplitude of total reflections; (v) two sonar parameters are renderized: Echo intensity and pulse distance. On CPU domain, the shader data is sorted in beam parts, where: (vi) a distance histogram correlates the pixels with respective bins; (vii) the bin intensity is computed by energy normalization; (viii) noise simulation degrades the sonar data; (ix) the noisy bin intensity is stored as a sonar data structure on Rock.}
	\label{fig:sonar_sim}
\end{sidewaysfigure}

The pipeline of the proposed sonar simulator is depicted in Fig. \ref{fig:sonar_sim}, and bridges two domains. On \textbf{GPU domain}, the engine computes reflections using an approach based on a selective rasterized ray-tracing. The resulting sonar rendering parameters are processed into the simulated acoustic data on \textbf{CPU domain}, where the sonar image is presented. This approach is detailed into the following subsections.

% =======================================================================================

% - UNDERWATER SCENE: integration between Rock, Gazebo and OpenSceneGraph
\subsection{Representation of underwater scene}
\label{dev:uwscene}

Underwater environment is defined with Rock-Gazebo integration \citep{watanabe2015}. Gazebo handles with physical simulations, where the hydrostatic and hydrodynamic forces and moments are modelled and applied on underwater vehicles, and provides access to the simulated objects and data; osgOcean, a plugin for OpenSceneGraph, renders the ocean with several visual effects such as sunlight, ocean surface foam, water turbidity, and light absorption and scattering. Rock framework manages the communication and synchronization between simulated components and displays the virtual scenario. Environment model and robot parts are described by SDF files.

% =======================================================================================

% - SONAR DEVICE: specialized camera with proper FOV, depth of field and resolution
\subsection{Sonar rendering on GPU}
\label{dev:rendering}

A virtual camera, properly configured with the desired \textbf{sonar settings} (\textit{i.e.}, pose, field of view, range and resolution), samples the \textbf{underwater simulated scene} (Fig. \ref{fig:sonar_sim} (i)). In vertex and fragment shaders, the captured rendering area passes by a rasterization and selective ray-tracing scheme, where the deferred shading provides the information needed to compute the primary reflections and only reflected areas are ray-traced for secondary reflections. This effectively enables a multipath propagation, prevents a whole interaction of intersection tests, and produces the same result in comparison of a full ray-tracer.

% =======================================================================================

% - SONAR PARAMETERS: deferred shading to compute the distance and normal values from captured scene
%     + PRIMARY REFLECTIONS - RASTERIZATION:
%         * Normal mapping / TBN matrix
%         * Surface reflectance
\subsubsection{Rasterized reflections}
\label{sim:rasterized}

The first reflection comes from the closest intersection of source wave with a scene object in 3D space. In order to improve the performance of findings the closest intersections, this work uses the deferred shading technique to mimic the first reflections with a sound wave. Rather than launching individual rays through the virtual environment, the \textbf{primary reflections} take advantage of two geometric information stored in G-buffers (position and normal vectors in world space) to compute the sonar rendering parameters during rasterization process (Figure \ref{fig:sonar_sim} (ii)):

\begin{itemize}
	\item \textit{Pulse distance}: Reproduces the length of sound wave. This parameter uses the depth information to compute the Euclidean distance between camera center and object surface, as defined by $r$ in Eq. (\ref{eq:geom:b}).
	\item \textit{Echo intensity}: Simulates the backscattered power of sound wave. The value is initially obtained from the normal incidence concerning the virtual camera.
\end{itemize}

Multiple factors can affect the strength of the reflected sound waves. In order to produce more realistic sonar images, four phenomena are considered here: Surface irregularities, material reflectance, sound attenuation and speckle noise. The former property enables the Lambertian diffuse reflection by applying normal map, an RGB texture, which changes the normal directions and, as a consequence, fakes roughness at the object surface with no additional polygons. The material reflectance deals with the acoustic reflectivity of sound waves, whose echoes are stronger from objects with densities different than water. So rocks, air-filled objects and compact gas reflect more sonar energy than softer surface types, like plastic and mud \citep{christ2013}. In this context, when an object has the reflectivity defined, this value is multiplied by the echo intensity. These two characteristics are detailed in \cite{cerqueira2017}. Next the influence of sound attenuation and speckle noise effects are presented.

% =======================================================================================

%     + SECONDARY REFLECTIONS - ray-tracing:
%         * Split the scene objects in triangles (vertices position, normal and centroid)
%         * Pass the triangles data as texture to shader
%         * Compute the world origin point + direction from the rasterization
%         * Selective rays: Perform rays only for normal > 0
%         * Optimized ray-triangle intersection:
%             a. Test ray-box intersection to check if the ray intersects an object
%             b. If a. yes, test the ray-triangle intersection
%             c. If b. yes, store the distance and normal values for secondary reflection
\subsubsection{Selective ray-traced reflections}
\label{sim:raytraced}

% ray-tracing + primary and second reflections + summary of ray tracer
Ray-tracing extends the wave propagation theory to simulate effects like reflections, ambient occlusions and shadows, but at a great computational cost. For highly complex scenes, this model generally becomes time-consuming due to excessive amount of intersection tests. In this work, the world position and normal vectors from G-buffers are used to compute the primary reflections of sound wave through the virtual scene, which identify where each ray starts and in which direction it should be reflected. Then the proposed ray-tracer starts the \textbf{secondary reflections} by selecting all pixels with surface normal values greater than zero to be traced (Fig. \ref{fig:sonar_sim} (iii)). In practice, this scheme propagates few rays when compared to a full ray-tracer, resulting in a significant speed-up with no significant loss of  information.

% Any surface can be approximated by triangulation
Testing if a ray intersects any surface requires analyzing all objects in the scene. According to the complexity of geometric surfaces, these scene objects can be described by simple shapes like spheres, cylinders and planes, or using mathematical models such as polygon meshes and splines for high detailed representations. In this context, ray-geometry intersection methods have to deal with each supported type of surface, increasing then the complexity of the implementation, drastically. Here all objects in the virtual underwater scene are depicted as triangulated meshes by using tessellation at rendering time, and the triangles data (\textit{i.e.}, vertices, surface normal, and centroid) feed the shader as textures. This way, any arbitrary surface can be rendered since each camera ray is tested against every individual triangle producing each polygon object in the scene with ray-triangle intersections.

The amount of time to compute ray-triangle intersections is directly proportional to the number of triangles in the scene. The rendering time can be saved by reducing the number of intersection tests \citep{akenine2018}. The selective ray-tracer is accelerated by bounding volume and a classic axis-aligned bounding boxes (AABBs) algorithm \citep{williams2005}, as follows: For each object in the scene, one box encapsulates all vertices; if the ray does not intersect a box, it is not able to intersect any triangle within this bounding volume; otherwise, the ray is tested against each triangle contained into the box with a Möller-Trumbore intersection algorithm \citep{moller1997}; in case of new intersection, the pulse distance and echo intensity values between triangle and ray origin are stored in a resulting image. This approach reproduces the secondary reflections by saving a significant number of calls to the ray-triangle routine, being summarized in the Algorithm \ref{alg:2ndreflections}.

% Insert the algorithm
\begin{algorithm}[t]
	\scriptsize
	\caption{Selective ray-tracer in GPU}
	\label{alg:2ndreflections}
	\begin{algorithmic}[1]
	\Function{SecondaryReflections}{$first$}
		\State $second \gets (0, 0)$
		\ForAll{$n$ \textbf{in} $first.normals$ \textbf{such that} $n$\textgreater~$0$}
			\State $[orig,~dir] \gets GetWorldCoordinates(n)$
			\State $ray \gets CalculateRay(orig,~dir)$
			\ForAll{$box$ \textbf{in} $boxes$}
				\State $intersection \gets IntersectBox$($ray$,$box$)
				\If{$intersection.hit$}
					\ForAll{$triangle$ \textbf{in} $box$}
						\State $intersection \gets IntersectTriangle$($ray$,$triangle$)
						\If{$intersection.hit$}
							\State $normal \gets triangle.normal$
							\State $distance \gets Length$($ray$,$triangle$)
							\State $second$.$writeReflection$($normal$,$distance$)
						\EndIf
					\EndFor
				\EndIf
			\EndFor
		\EndFor
	\EndFunction
	\end{algorithmic}
	\end{algorithm}

% =======================================================================================

%     + UNIFIED
%         * Unify the primary and secondary reflections in one shader image
%         * Apply sound attenuation
%         * Outputs the shader image
% include the relationship between alpha and intensity
% merge into one shader image + sound attenuation
\subsubsection{Unified reflections}
\label{dev:unified}

After the computation of primary and secondary reflections, the corresponding results are blended in an unified shader image with echo intensity and pulse distance values, and finally the \textbf{signal attenuation} effect is applied (Fig. \ref{fig:sonar_sim} (iv)). Since the water is a dissipative medium, the sound intensity decreases exponentially with the distance travelled by absorption and spreading, while propagating. Equation (\ref{eq:atten:a}) expresses the attenuation coefficient $\alpha$, which can be converted to Np/km, as follows:

\begin{equation}
	\label{eq:atten:e}
	1 dB = \frac{1}{20 \log_{10}e}Np \approx 0.0115 Np, \quad \gamma = 0.0115 \alpha \, .
\end{equation}

The sound pressure decays according to

\begin{equation}
	\label{eq:atten:f}
	p_{d} = p_{0}e^{-\gamma d} \, .
\end{equation}

Within the same medium, the sound intensity is proportional to the average of the squared pressure \citep{dunn2015}

\begin{equation}
	\label{eq:atten:g}
	I \approx p^{2} \, .
\end{equation}

\begin{figure}[!t]
	\centering
	\begin{subfigure}[b]{0.4\linewidth}
		\includegraphics[width=\textwidth,height=45mm]{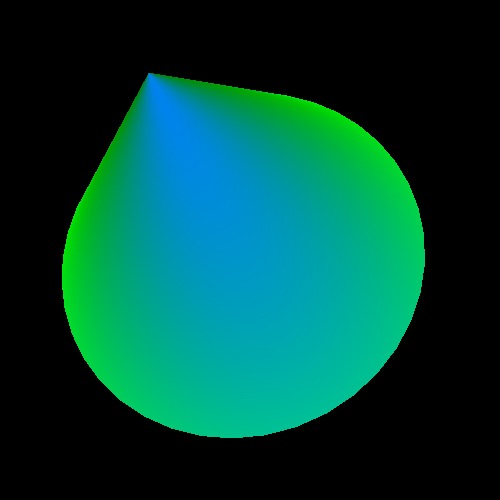}
		\caption{}
    \label{fig:atten:shader_without}
	\end{subfigure}
	\begin{subfigure}[b]{0.45\linewidth}
    \includegraphics[width=\textwidth,height=45mm]{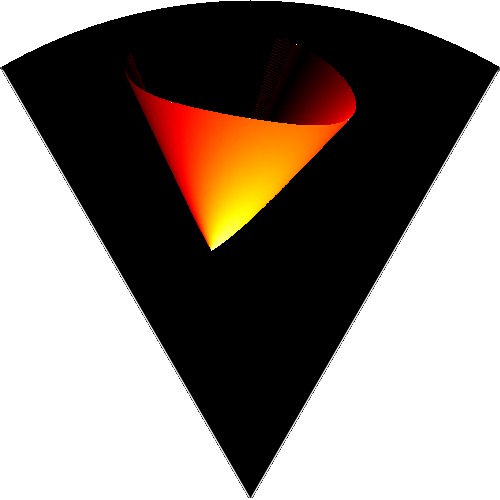}
		\caption{}
    \label{fig:atten:sonar_without}
	\end{subfigure}
	\begin{subfigure}[b]{0.4\linewidth}
    \includegraphics[width=\textwidth,height=45mm]{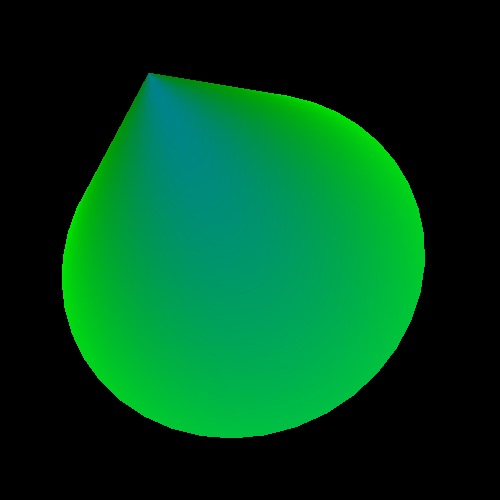}
		\caption{}
    \label{fig:atten:shader_with}
	\end{subfigure}
	\begin{subfigure}[b]{0.45\linewidth}
    \includegraphics[width=\textwidth,height=45mm]{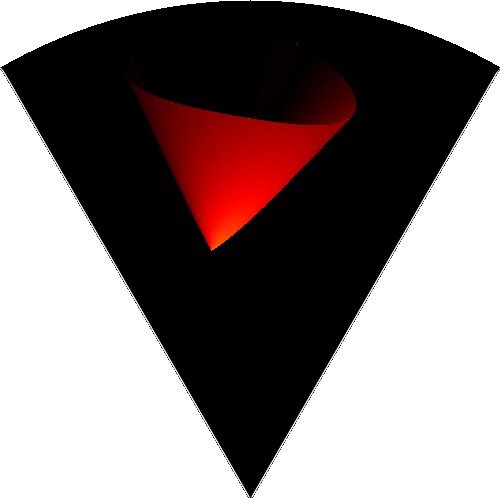}
		\caption{}
    \label{fig:atten:sonar_with}
	\end{subfigure}

	\caption{Example of different attenuation coefficient values, $\alpha$, applied on scene rendering of a cone. In shader images, the blue and green channels express the pulse distance and echo intensity data, respectively, for \subref{fig:atten:shader_without} $\alpha = 0$ dB/km and \subref{fig:atten:shader_with} $\alpha = 0.013$ dB/km. The corresponding sonar images are depicted in (\subref{fig:atten:sonar_without}) and \subref{fig:atten:sonar_with}. By applying sound attenuation effect, the echo intensity decreases exponentially with distance from the source.}
  \label{fig:atten}
	\end{figure}

Therefore

\begin{equation}
	\label{eq:atten:h}
	I_{d} = I_{0}e^{-2\gamma d} \, ,
\end{equation}
where the initial intensity $I_{0}$ is reduced to $I_{d}$ at a distance $d$ (in km), and the attenuation coefficient $\gamma$ in Np/km. An example of the effect is showed in Fig. \ref{fig:atten}, where the attenuation coefficient weakens the acoustic intensity with increasing propagation distance.

The final pulse distance and echo intensity values are organized as blue and green channels of shader image, respectively (see \textbf{sonar rendering parameters} in Fig. \ref{fig:sonar_sim}(v)). These values range from 0 to 1. For the echo intensity, zero means no energy, while one denotes maximum reflection returned. For the pulse distance, the minimal and maximum values express the near and far planes, respectively. At the end, the sonar parameters are rendered to a floating-point RGBA texture, using a framebuffer object (FBO), to avoid loss of precision mainly for the pulse distance values.

% =======================================================================================

% - CONVERT TO SONAR DATA
%     + Split the shader image in beam parts, according to number of beams and bearing data
% 	+ For each beam, group the pixels in bins by depth histogram
% 	+ For each bin, compute intensity by energy normalization
% 	+ Apply speckle noise (multiplicative and additive components)
% 	+ Store the data into vector
% - DISPLAY SONAR IMAGE
% 	+ Polar x Cartesian representation
\subsection{Generating the sonar image on CPU}
\label{dev:simulation}

On CPU domain, the resulting sonar rendering parameters are converted into the respective acoustic data. While the azimuth angle is radially spaced over the virtual camera, the elevation angle is lost during sonar projection geometry. This process implies all pixels belong a column have the same bearing angle. The shader image columns are then divided into a number of beam parts. For each beam section, a \textbf{distance histogram} groups the pixels in bins, according to pulse distance values (Fig. \ref{fig:sonar_sim} (vi)). Finally, the accumulated bin intensity value, $I$, is computed with an \textbf{energy normalization} function (Fig. \ref{fig:sonar_sim} (vii)), given by

\begin{equation}
    \label{eq:energy_norm}
    I(r,\theta) = \sum\limits_{x=1}^N \frac{1}{N} S(i_{x}),
\end{equation}
where $(r,\theta)$ are polar coordinates, $N$ is the number of pixels with respect to one bin, $x$ is the pixel index, and $S$ is a sigmoid function applied over the echo intensity $i_{x}$.

% speckle noise
Due to the acquisition process and complexity of underwater sound propagation, acoustic devices suffer from speckle noise and random variations of echo intensity. All these make further data interpretation difficult. To simulate the speckle noise in the resulting image, Eq. (\ref{eq:noise}) is used according to \textbf{noise simulation} functions (Fig. \ref{fig:sonar_sim} (viii)). The multiplicative component follows a non-uniform Gaussian distribution, while the additive one is denoted by a Gaussian random variable with zero mean and standard deviation $\sigma^{2}$. The noise model is repeated for each acoustic frame.

% acoustic representation
The simulation ends with the conversion of noisy intensity values into a \textbf{data structure of corresponding beam} (Fig. \ref{fig:sonar_sim}). The sonar data is latter displayed in Cartesian coordinates on Rock framework, according to Eq. (\ref{eq:geom:c}).

% =======================================================================================

\section{Experimental evaluation}
\label{results:intro}

\subsection{Setup}
\label{results:setup}

Our sonar simulator was implemented in C++, OpenCV and OpenSceneGraph on CPU. Shaders relies in massive parallelism available on GPU to render the underwater scene using rasterization and ray-tracing, and Ruby scripts connect and monitor components on Rock framework. All experiments were performed on an Intel Core i7-8750H 2.20 GHz, running with 16 GB DDR3 RAM, NVIDIA GeForce GTX 1060 video card and Ubuntu 16.04 64 bits operating system.

% =======================================================================================

\subsection{Visual quality}
\label{results:qualitative}

\begin{figure}[!t]
	\centering
	\addtolength{\leftskip} {-1cm}
	\addtolength{\rightskip}{-1cm}
	\begin{subfigure}[b]{0.525\linewidth}
		\includegraphics[width=\textwidth,height=50mm]{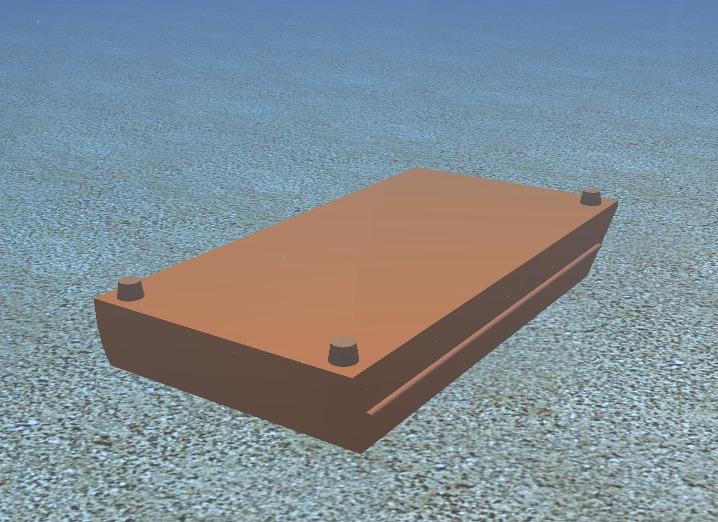}
		\caption{}
    \label{fig:ferry_photo_sim}
	\end{subfigure}
	\begin{subfigure}[b]{0.525\linewidth}
		\includegraphics[width=\textwidth,height=50mm]{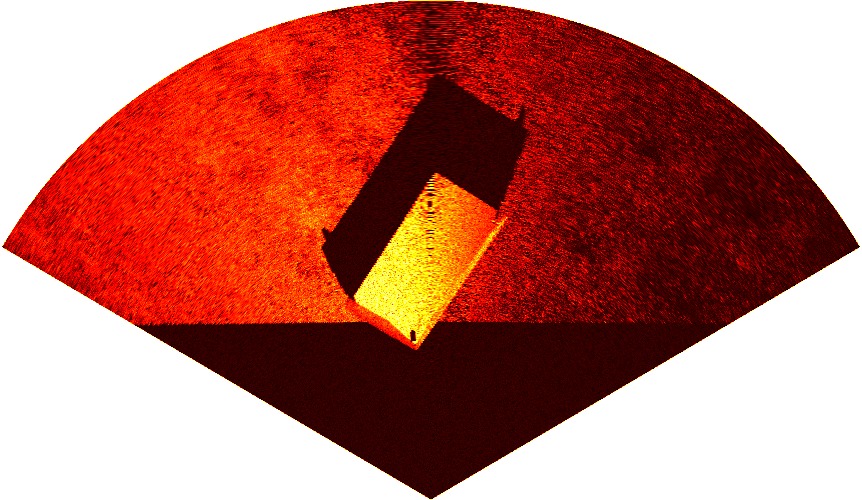}
		\caption{}
		\label{fig:ferry_cart_sim}
	\end{subfigure}
	\begin{subfigure}[b]{0.525\linewidth}
    \includegraphics[width=\textwidth,height=50mm]{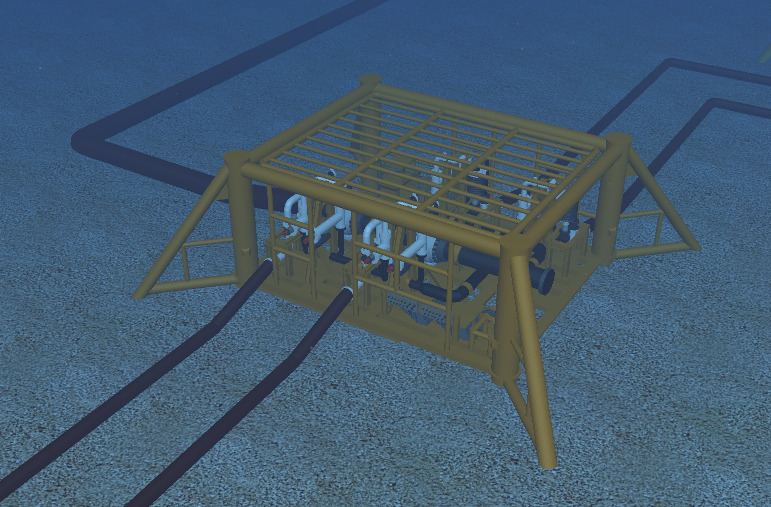}
		\caption{}
		\label{fig:cooler_photo_sim}
	\end{subfigure}
	\begin{subfigure}[b]{0.525\linewidth}
    \includegraphics[width=\textwidth,height=50mm]{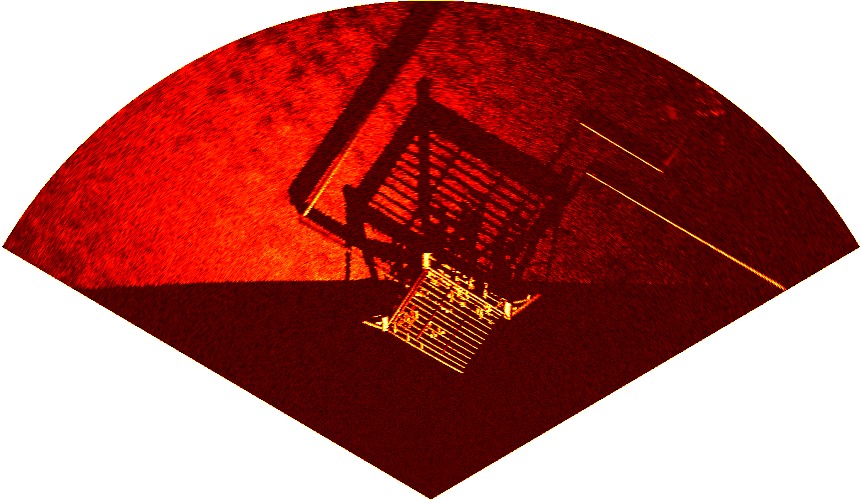}
		\caption{}
		\label{fig:cooler_cart_sim}
	\end{subfigure}
	\caption{Demonstration of acoustic images generated by the sonar simulation system: \subref{fig:ferry_photo_sim} and \subref{fig:cooler_photo_sim} are the insonified targets in the underwater environment; \subref{fig:ferry_cart_sim} and \subref{fig:cooler_cart_sim} present sonar data produced by FLS device. The simulated representation of wrecked ferry in Fig. \ref{fig:view_representation} is depicted by \subref{fig:ferry_photo_sim} and \subref{fig:ferry_cart_sim}.}
	\label{fig:results:fls}
	\end{figure}

\begin{figure}[!t]
	\centering
	\addtolength{\leftskip} {-1cm}
	\addtolength{\rightskip}{-1cm}
	\begin{subfigure}[b]{0.525\linewidth}
		\centering
		\includegraphics[width=\textwidth,height=50mm]{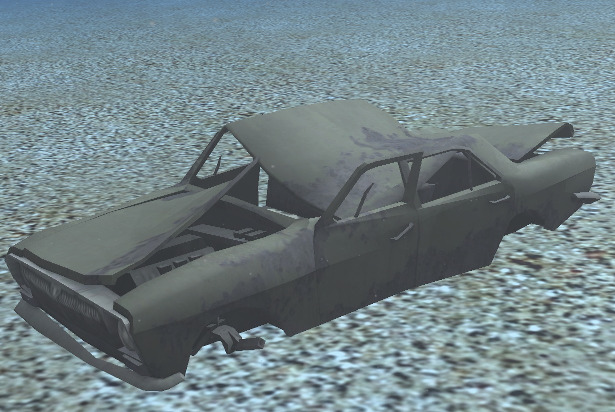}
		\caption{}
		\label{fig:car_photo_sim}
	\end{subfigure}
	\begin{subfigure}[b]{0.525\linewidth}
		\centering
		\includegraphics[width=0.85\textwidth,height=50mm]{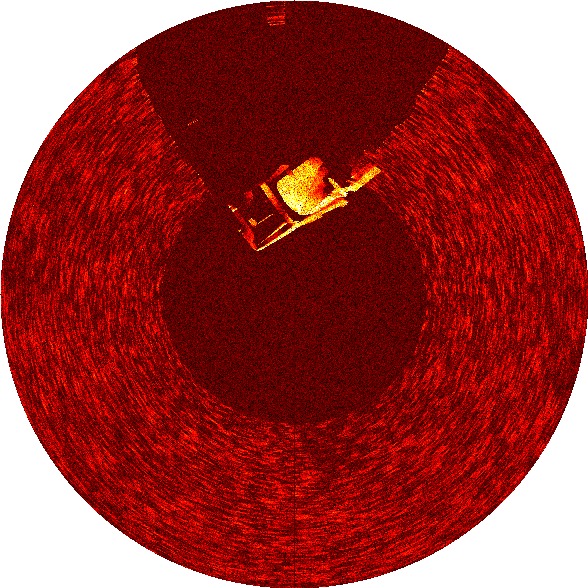}
		\caption{}
		\label{fig:car_cart_sim}
	\end{subfigure}
	\begin{subfigure}[b]{0.525\linewidth}
		\includegraphics[width=\textwidth,height=50mm]{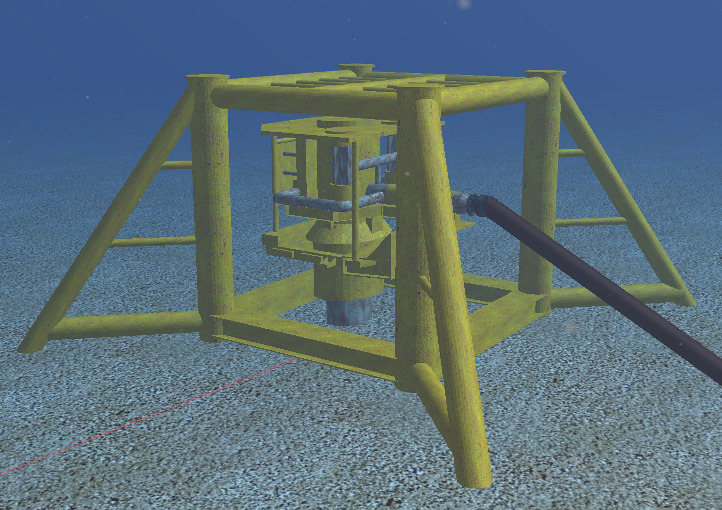}
		\caption{}
		\label{fig:xmastree_photo_sim}
	\end{subfigure}
	\begin{subfigure}[b]{0.525\linewidth}
		\centering
		\includegraphics[width=0.85\textwidth,height=50mm]{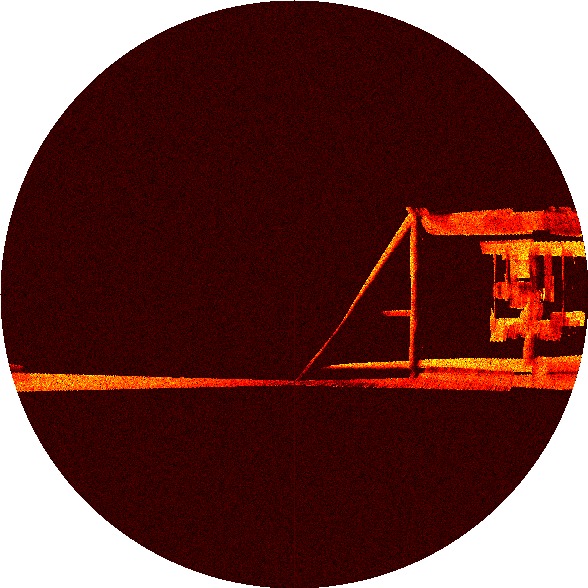}
		\caption{}
		\label{fig:xmastree_cart_sim}
	\end{subfigure}
	\caption{Demonstration of acoustic images generated by the sonar simulation system: \subref{fig:car_photo_sim} and \subref{fig:xmastree_photo_sim} are the insonified targets in the underwater environment; \subref{fig:car_cart_sim} and \subref{fig:xmastree_cart_sim} denotes the virtual acoustic representations by MSIS device mounted in horizontal and vertical orientations, respectively.}
	\label{fig:results:msis}
\end{figure}

To evaluate the visual quality of the images generated by the simulator, FLS and MSIS devices, equipping a virtual AUV, were simulated to insonify four different scenarios. In the first scene, illustrated in Fig. \ref{fig:ferry_photo_sim}, a wrecked ferry was used; the shape of the ferry is insonified in the FLS image, as well as the corrugated seabed after a normal mapping technique, as can be seen in the sonar chart of Fig. \ref{fig:ferry_cart_sim}; given the material reflectance is defined, the target is distinguishable from other scene components. The second scene consists of a subsea cooler connected to pipelines in an oil production field (see Fig. \ref{fig:cooler_photo_sim}); front faces of targets and the shadows occluding part of the scene are clearly visible in FLS chart image (see Fig. \ref{fig:cooler_cart_sim}); the echo intensity of acoustic image is perturbed with speckle noise pattern; also, the attenuation effect weakens the intensity of the farthest bins from sonar head. The third scene contains a destroyed car on the seafloor, and is depicted in Fig. \ref{fig:car_photo_sim}; using the MSIS in horizontal orientation, the regions with approximated perpendicular angle to the sonar viewpoint, or multiple returns, are identified as brightest areas in the sonar chart of Fig. \ref{fig:car_cart_sim}; the image of this sonar chart is also characterized by the granular disturbance of the speckle noise. An offshore Christmas tree is the main target of the last scene (see Fig. \ref{fig:xmastree_photo_sim}); an MSIS vertically mounted on the AUV captures the slice scanning of seafloor and the Christmas tree (see Fig. \ref{fig:xmastree_cart_sim}).

For all experiments, the initial bins presented low intensity values, caused by the lack of acoustic feedback in short ranges. The rendering of complex scenes was also addressed, highlighting the details present on geometries of insonified objects. Yet, acoustic shadows contain valuable information for the accurate interpretation of the sonar images; depending on the angle of incidence, the shadows can present more details than the sonar acoustic return, as illustrated by the pipelines in Fig. \ref{fig:cooler_cart_sim}.

% =======================================================================================

\subsection{Computational time}
\label{results:performance}

\begin{sidewaystable}
	\centering
	\scriptsize
	\caption{Processing time to generate 500 samples from FLS sensor with different configurations.}
	\begin{tabular}{@{}cccccccccccc@{}}
		\toprule
		\multicolumn{3}{c}{Setup} && \multicolumn{3}{c}{\cite{cerqueira2017}} && \multicolumn{3}{c}{Ours}\\
		\cmidrule{1-3} \cmidrule{5-7} \cmidrule{9-11}
		\# of beams	&	\# of bins	&	Field of view (w x h)	&& Avg. time (ms)	& Std dev (ms)	& Frame rate (fps) && Avg. time (ms)	& Std dev (ms)	& Frame rate (fps)\\
		\cmidrule{1-3} \cmidrule{5-7} \cmidrule{9-11}
		128	& 500	&	$120^{\circ}$ x $20^{\circ}$	&& 49.0    	& 1.3	& 20.5	&& 86.1    & 12.5	& 11.7	\\
		128	& 1000	&	$120^{\circ}$ x $20^{\circ}$	&& 95.4    	& 1.7	& 10.5	&& 132.3   & 16.5	& 7.6	\\
		256	& 500	&	$120^{\circ}$ x $20^{\circ}$	&& 59.3   	& 1.6	& 16.9	&& 101.7   & 14.5	& 9.9	\\
		256	& 1000	&	$120^{\circ}$ x $20^{\circ}$	&& 117.2   	& 2.4	& 8.6	&& 155.8   & 15.1	& 6.5	\\
		128	& 500	&	$90^{\circ}$ x $15^{\circ}$		&& 56.4		& 2.5	& 17.8	&& 97.4    & 12.3	& 10.3	\\
		128	& 1000	&	$90^{\circ}$ x $15^{\circ}$		&& 118.6	& 2.5	& 8.5	&& 144.4   & 12.7	& 7.0	\\
		256	& 500	&	$90^{\circ}$ x $15^{\circ}$		&& 66.7     & 1.3	& 15.0	&& 111.7   & 13.0	& 9.0	\\
		256	& 1000	&	$90^{\circ}$ x $15^{\circ}$		&& 137.5    & 5.8	& 7.3	&& 169.3   & 16.0	& 6.0	\\
	\bottomrule
	\label{tab:time:fls}
	\end{tabular}

	\vspace{2\baselineskip}

	\caption{Processing time to generate 500 samples from MSIS sensor with different configurations.}
	\begin{tabular}{@{}cccccccccccc@{}}
		\toprule
		\multicolumn{2}{c}{Setup} && \multicolumn{3}{c}{\cite{cerqueira2017}} && \multicolumn{3}{c}{Ours}\\
		\cmidrule{1-2} \cmidrule{4-6} \cmidrule{8-10}
		\# of bins	&	Field of view (w x h)	&& Avg. time (ms)	& Std dev (ms)	& Frame rate (fps) && Avg. time (ms)	& Std dev (ms)	& Frame rate (fps)\\
		500		&	$3^{\circ}$ x $35^{\circ}$	&& 28.4    	& 2.1	& 35.4	&& 62.8    & 6.8	& 16.0	\\
		1000	&	$3^{\circ}$ x $35^{\circ}$	&& 34.8    	& 3.1	& 28.8	&& 70.0    & 6.9	& 14.3	\\
		500		&	$2^{\circ}$ x $20^{\circ}$	&& 30.7   	& 3.0	& 32.6	&& 69.1    & 5.6	& 14.5	\\
		1000	&	$2^{\circ}$ x $20^{\circ}$	&& 37.7   	& 3.6	& 26.6	&& 77.2    & 6.9	& 13.0	\\
		\bottomrule
		\cmidrule{1-2} \cmidrule{4-6} \cmidrule{8-10}
	\end{tabular}
	\label{tab:time:msis}

\end{sidewaystable}

To evaluate the computational cost of our simulator, we built a data set containing four different geometric shapes randomly positioned along the sonar viewport, for each frame: Cylinder, box, sphere and cone. These geometric shapes provides a good variation in the amount of triangle meshes during tesselation process. To measure the execution time, three metrics were used, as such: Average time, standard deviation and frame rate. For each iteration, the simulator executed the same series of tasks: 1) read input scene frame and sonar device settings; 2) solve acoustic model equations; 3) output virtual sonar image. The data set was used to compare the performance of the sonar simulator with the method proposed in \cite{cerqueira2017}, with the same hardware resources. The results are summarized in Tables \ref{tab:time:fls} and \ref{tab:time:msis}.

According to the results, our simulator spends almost double the time than the method in \citep{cerqueira2017}; this can be explained by the fact that our simulator constructs the scene by ray-tracing the secondary reflections, turning the tessellation of shapes and additional acoustic phenomena to produce an extra computational cost. Yet, since the processing time of ray-tracing depends on the reflective areas presented in the camera viewport of the simulator (determined by the scene to be rendered), the standard deviation was much greater than that based on pure rasterization. On the other hand, the parallel approach on GPU and the significant reduction of ray-triangle intersection calls retained our approach close to the operation of real sonar devices. According to the list of main FLS available in the market \citep{hurtos2014}, the Tritech Gemini 720i sensor, with a field of view of 120\degree x 20\degree and 256 beams, owns refresh rates of 5-30 fps (range dependent).

In comparison with other simulators, for the FLS device, our rates are superior to the rates listed by \cite{demarco2015} (3 fps), \cite{mai2018} (1 fps) and \cite{sac2015} (2.5 min), even with additional acoustic phenomena present in the simulated sonar image. For MSIS type, to the best of our knowledge, there is no previous work with rates for comparison.

The number of bins and beams also impacts on the simulation performance. The former is directly proportional to image resolution; the amount of pixels to be processed increases with the number of bins. The latter determines the number of beam sections of shader images to be rendered.

% =======================================================================================

\subsection{Quantitative evaluation}
\label{results:quantitative}

\begin{figure}[!t]
	\centering
	\addtolength{\leftskip} {-2cm}
	\addtolength{\rightskip}{-2cm}
	\begin{subfigure}[b]{0.55\linewidth}
		\includegraphics[width=\textwidth,height=50mm]{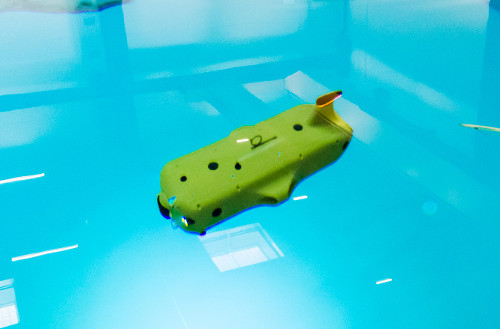}
		\caption{}
    \label{fig:flatfish_tests_dfki}
	\end{subfigure}
	\begin{subfigure}[b]{0.55\linewidth}
		\includegraphics[width=\textwidth,height=50mm]{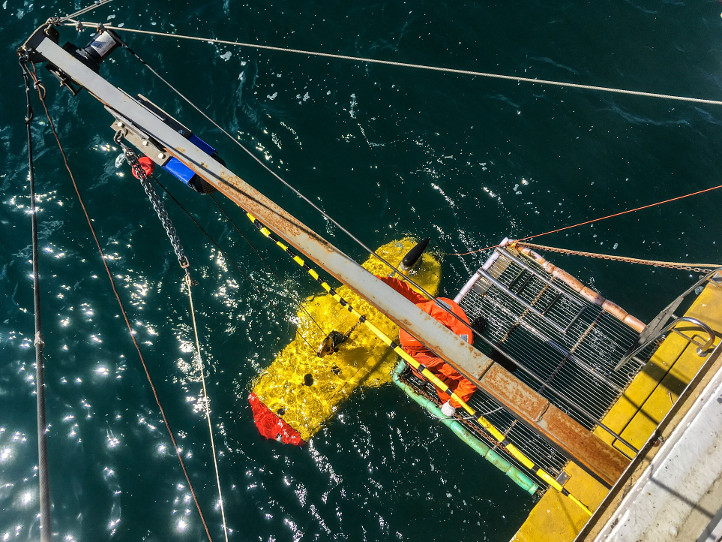}
		\caption{}
		\label{fig:flatfish_tests_salvador}
	\end{subfigure}

	\caption{Trials with FlatFish AUV during acoustic data acquisition for quantitative evaluation: \subref{fig:flatfish_tests_dfki} At DFKI RIC, Bremen, Germany; \subref{fig:flatfish_tests_salvador} In Todos os Santos Bay, Salvador, Brazil.}
	\label{fig:flatfish_tests}
\end{figure}

For quantitative analysis, two different scenarios were sampled by real FLS and MSIS sensors equipped on FlatFish AUV \citep{albiez2015}. In the former scenario, a Tritech Gemini 720i insonified a subsea safety isolation valve (SSIV) mockup on the seabed in Todos os Santos Bay, Salvador, Brazil; the latter scenario is comprised of a Tritech Micron DST sonar horizontally mounted to capture the tank walls surrounding the AUV at DFKI RIC, Bremen, Germany. Figure \ref{fig:flatfish_tests} present the FlatFish AUV during these trials. The aforementioned experiments were repeated in the virtual underwater scenario with the same targets, and our sonar system generated the corresponding acoustic representations. A summary of the experiments is depicted in Figs. \ref{fig:results:fls} and \ref{fig:results:msis}.

Similarity measurements between real and simulated sonar images depend on device configuration, environment characteristics, observable objects and acquisition viewpoint. Four metrics were chosen to compute the similarity between the acoustic frames of the real and simulated images: Mean-squared error (MSE), peak signal-to-noise (PSNR), structural similarity index measure (SSIM) \citep{yang2008} and multi-scale structural similarity index measure (MS-SSIM) \citep{wang2003}. In order to preserve the original data, polar-coordinate based images were used in this evaluation. Table \ref{tab:results:quant} summarizes the results found in comparison with the method proposed in \citep{cerqueira2017}. Values in the table were normalized to zero representing minimum similarity, while one denotes maximum correlation.

\begin{table*}[!t]
	\centering
	\addtolength{\leftskip} {-2cm}
	\addtolength{\rightskip}{-2cm}
	\scriptsize
	\caption{Similarity evaluation results between real and simulated sonar images.}
	\begin{tabular}{@{}cccccccccccc@{}}
		\toprule
		\multicolumn{2}{c}{Scene} && \multicolumn{4}{c}{\cite{cerqueira2017}} && \multicolumn{4}{c}{Ours}\\
		\cmidrule{1-2} \cmidrule{4-7} \cmidrule{9-12}
		Device 	& Target 	&& MSE & PSNR & SSIM & MS-SSIM		&& MSE & PSNR & SSIM & MS-SSIM \\
		\cmidrule{1-2} \cmidrule{4-7} \cmidrule{9-12}
		FLS 	& SSIV		&& 0.990 & 0.669 & 0.361 & 0.628	&& 0.994 & 0.690 & 0.405 & 0.683 \\
		MSIS	& Tank		&& 0.996 & 0.761 & 0.834 & 0.852	&& 0.996 & 0.760 & 0.832 & 0.849 \\
		\bottomrule
	\end{tabular}
	\label{tab:results:quant}
\end{table*}

\begin{figure}[!t]
	\centering

	\setcounter{subfigure}{0}
	\begin{subfigure}[b]{0.49\linewidth}
		\includegraphics[width=\textwidth,height=45mm]{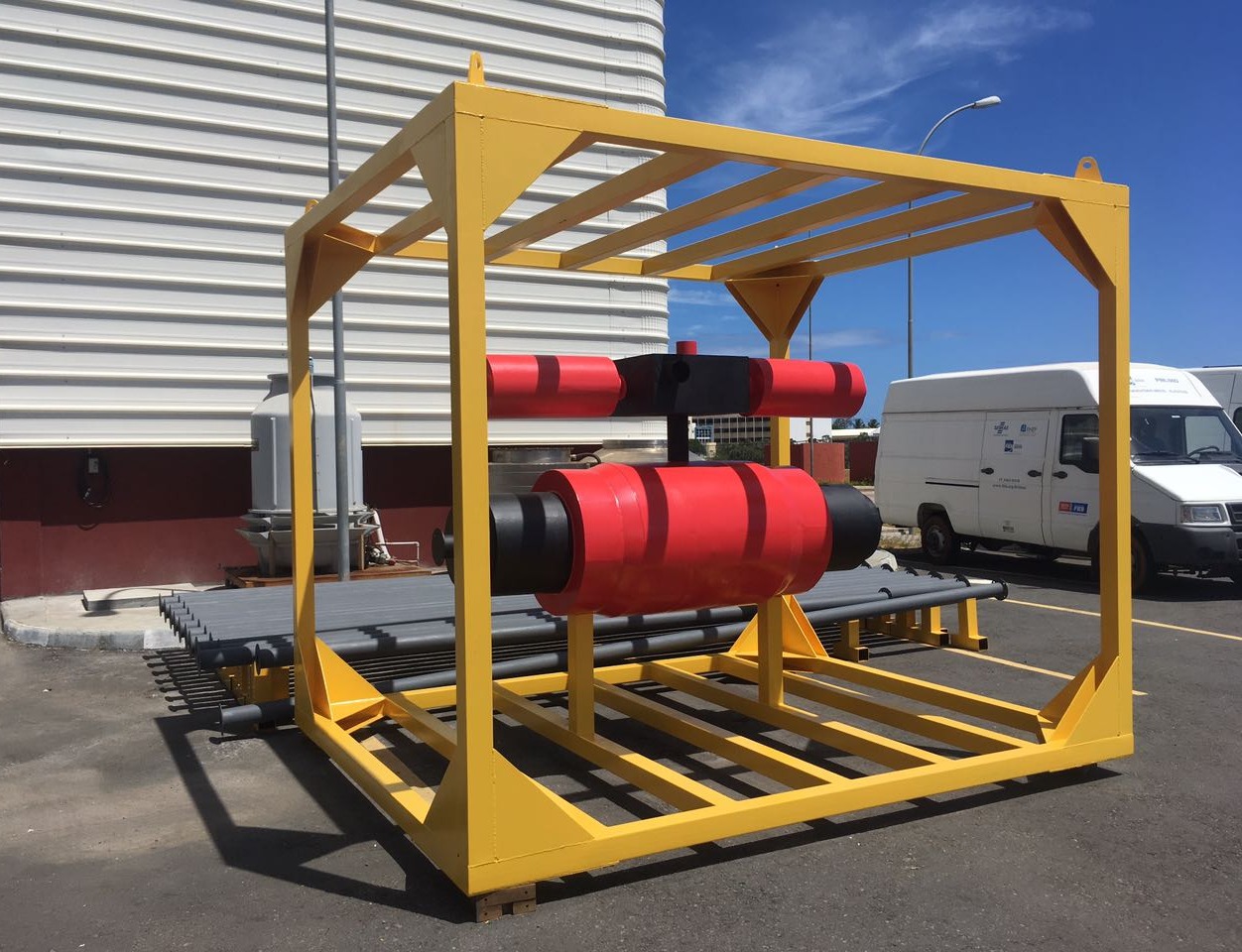}
		\caption{}
		\label{fig:ssiv_photo_real}
	\end{subfigure}
	\setcounter{subfigure}{2}
	\begin{subfigure}[b]{0.49\linewidth}
		\includegraphics[width=\textwidth,height=45mm]{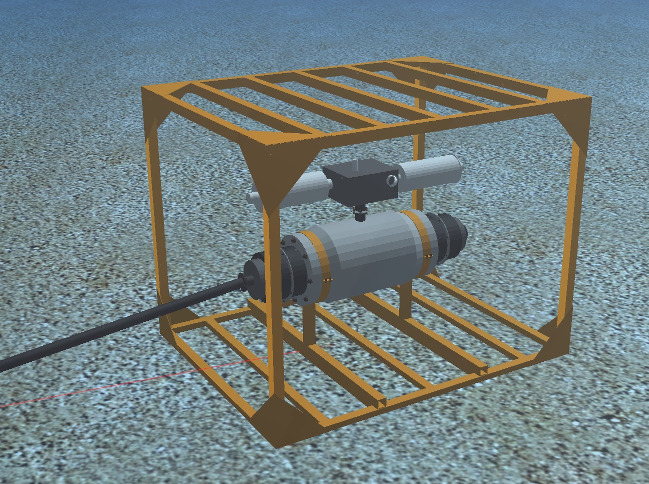}
		\caption{}
		\label{fig:ssiv_photo_sim}
	\end{subfigure}

	\setcounter{subfigure}{1}
	\begin{subfigure}[b]{0.49\linewidth}
		\includegraphics[width=\textwidth,height=45mm]{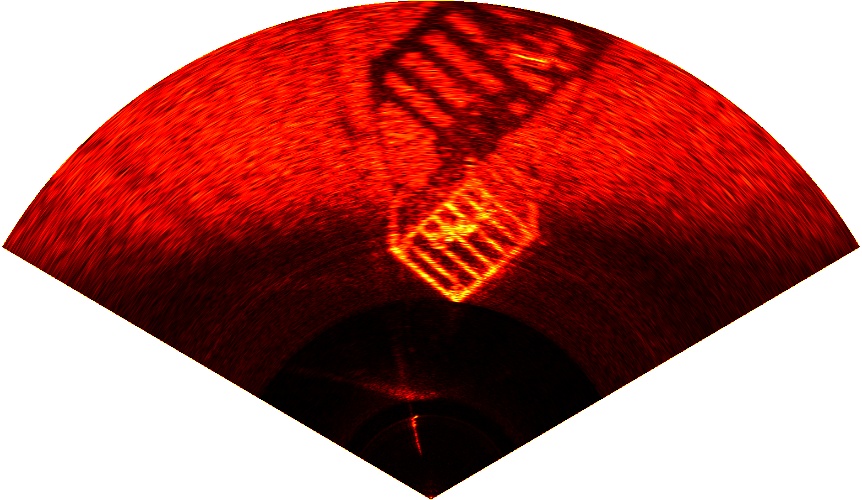}
		\caption{}
		\label{fig:ssiv_cart_real}
	\end{subfigure}
	\setcounter{subfigure}{3}
	\begin{subfigure}[b]{0.49\linewidth}
		\includegraphics[width=\textwidth,height=45mm]{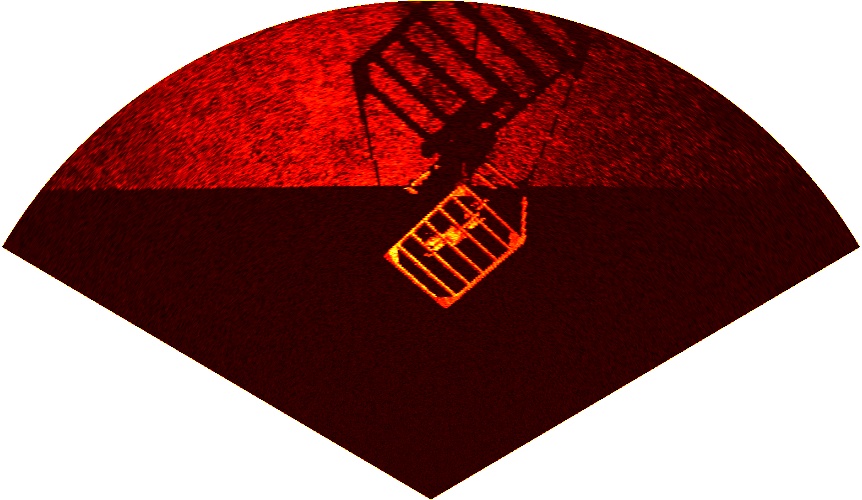}
		\caption{}
		\label{fig:ssiv_cart_sim_ours}
	\end{subfigure}

	\hfill
	\setcounter{subfigure}{4}
	\begin{subfigure}[b]{0.49\linewidth}
		\includegraphics[width=\textwidth,height=45mm]{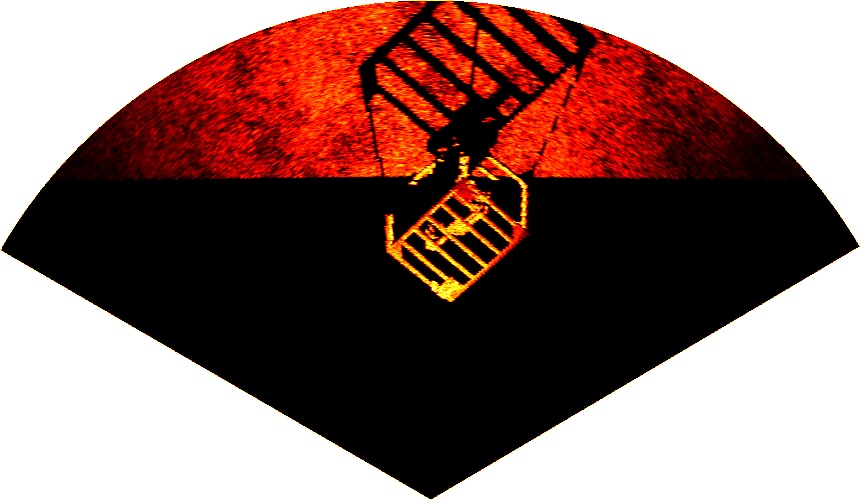}
		\caption{}
		\label{fig:ssiv_cart_sim_cerqueira2017}
	\end{subfigure}

	\caption{Experimental results with FLS device: \subref{fig:ssiv_photo_real} An SSIV mockup; \subref{fig:ssiv_cart_real} the acoustic data captured by Tritech Gemini 720i sonar equipped on the FlatFish AUV; \subref{fig:ssiv_photo_sim} a 3D model of the SSIV; \subref{fig:ssiv_cart_sim_ours} and \subref{fig:ssiv_cart_sim_cerqueira2017} are the acoustic data generated by our simulator and that one proposed in \cite{cerqueira2017}, respectively.}
	\label{fig:results:fls}
\end{figure}

In the FLS experiment, the values of our proposed work with MSE, PSNR, SSIM and MS-SSIM were higher to those values found in \citep{cerqueira2017}, mainly explained by the attenuation, additive noise and reverberation phenomena presented in the complex and full-detailed image. Conversely, SSIM individually presents lower performance for our proposed system, due to the sensitivity of this metric to changes on local intensity and contrast patterns on a very simple scene image (see Figs. \ref{fig:results:fls} \subref{fig:ssiv_cart_sim_ours} and \subref{fig:ssiv_cart_sim_cerqueira2017}). In MSIS experiment, MSE, PSNR, SSIM and MS-SSIM showed values approximately equal to the ones found in \cite{cerqueira2017}, what can be justified to the simplicity of the object edges insonified by the MSIS device. Indeed, the level of attenuation, speckle noise and reverberation was not enough in the image to define a gain on the simulated image in comparison with the real one, specifically regarding the quality of the image.

\begin{figure}[!t]
	\centering

	\setcounter{subfigure}{0}
	\begin{subfigure}[b]{0.49\linewidth}
		\includegraphics[width=\textwidth,height=45mm]{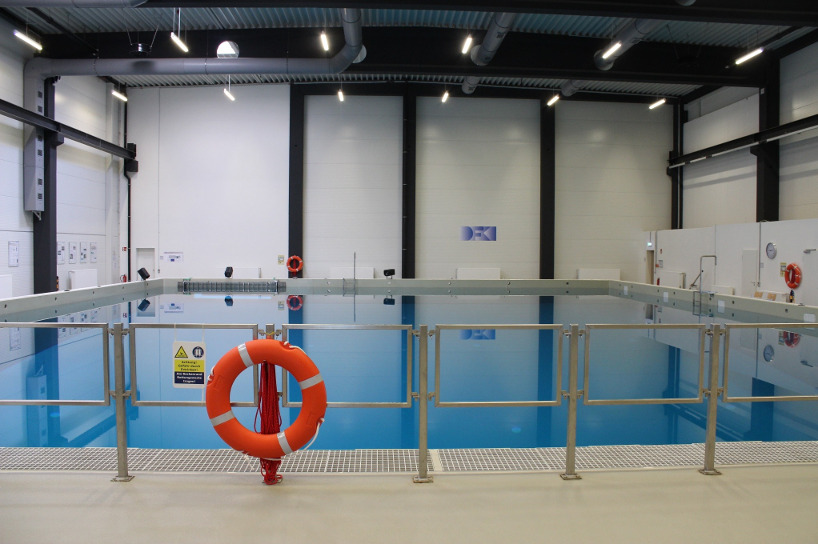}
		\caption{}
		\label{fig:tank_photo_real}
	\end{subfigure}
	\setcounter{subfigure}{2}
	\begin{subfigure}[b]{0.49\linewidth}
		\includegraphics[width=\textwidth,height=45mm]{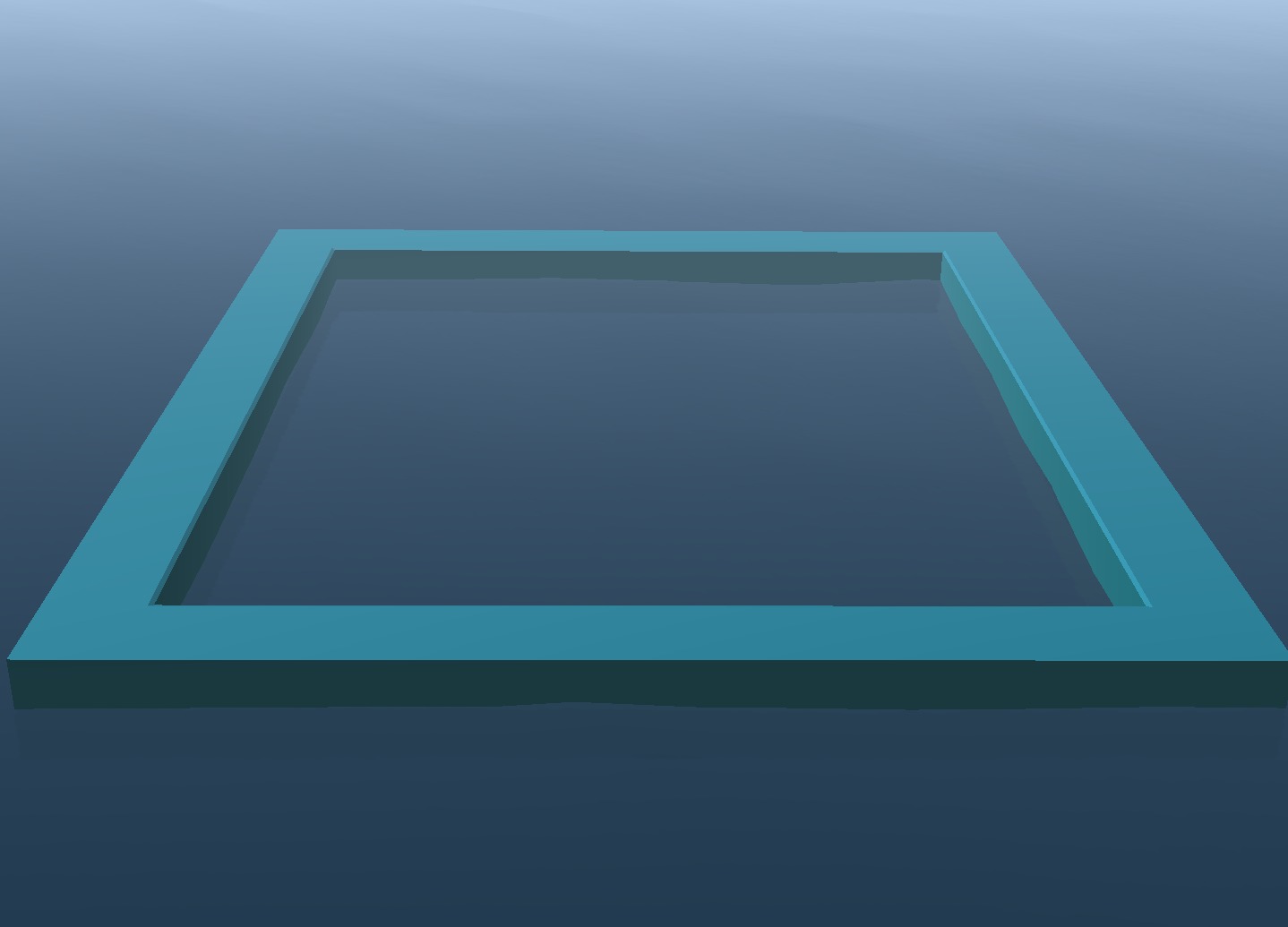}
		\caption{}
		\label{fig:tank_photo_sim}
	\end{subfigure}

	\setcounter{subfigure}{1}
	\begin{subfigure}[b]{0.49\linewidth}
		\centering
		\includegraphics[width=0.85\textwidth,height=45mm]{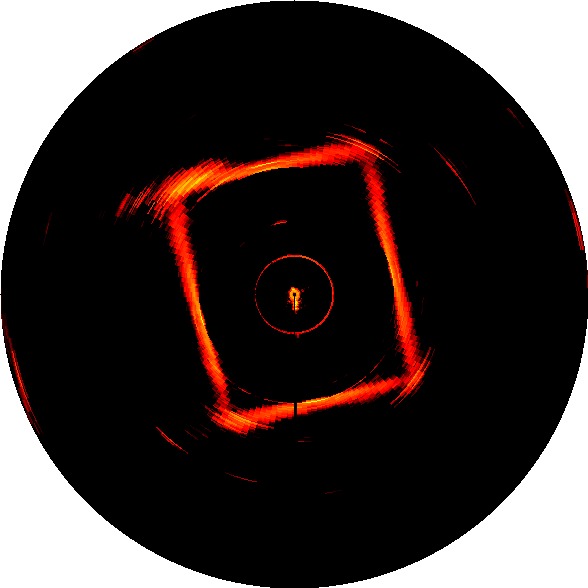}
		\caption{}
		\label{fig:tank_cart_real}
	\end{subfigure}
	\setcounter{subfigure}{3}
	\begin{subfigure}[b]{0.49\linewidth}
		\centering
		\includegraphics[width=0.85\textwidth,height=45mm]{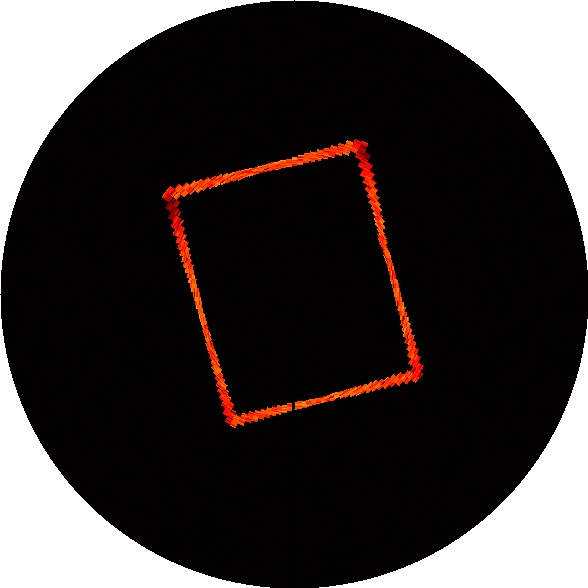}
		\caption{}
		\label{fig:tank_cart_sim_ours}
	\end{subfigure}

	\hfill
	\setcounter{subfigure}{4}
	\begin{subfigure}[b]{0.49\linewidth}
		\centering
		\includegraphics[width=0.85\textwidth,height=45mm]{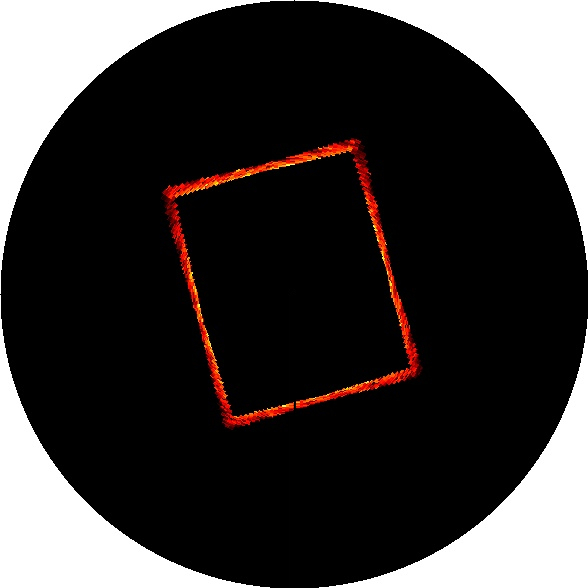}
		\caption{}
		\label{fig:tank_cart_sim_cerqueira2017}
	\end{subfigure}

	\caption{Experimental results with MSIS device: \subref{fig:tank_photo_real} DFKI tank; \subref{fig:tank_cart_real} the acoustic data captured by Tritech Micron DST sonar horizontally mounted on the FlatFish AUV; (\subref{fig:tank_photo_sim}) simulated tank; (\subref{fig:tank_cart_sim_ours}) and (\subref{fig:tank_cart_sim_cerqueira2017}) are acoustic data generated by our simulator and that one proposed in \cite{cerqueira2017}, respectively.}
	\label{fig:results:msis}
  \end{figure}

\section{Discussion and conclusions}
\label{conclusion}

% visual evaluation
Existing methods focus on simplified implementation of one specific sonar type, where the majority of underwater sound properties are disregarded. The proposed simulator here was able to reproduce the operation of two different sonar devices: FLS and MSIS. All experimental scenarios were defined to demonstrate phenomena usually found in real sonar images, such as speckle noise, transmission loss and material properties of insonified surfaces. It is noteworthy that the sea level is not considered during sonar rendering, turning this particular reverberation component absent from the computation of the final simulated image.

% computational time evaluation
The combination of rasterization and ray-tracing showed to speed up the overall sonar simulation time, in comparison with full ray-based techniques. This was achieved by reducing the number of launched rays, and at the same time not degrading the quality of the image (what was improved, in fact). The parallel ray-box routines on GPU also accelerated the intersection tests on ray-tracing algorithm. According to the results, the proposed simulator was able to compete with real sonar devices in terms of execution time, providing more realistic scenes than those generated by state-of-the-art methods.

% quantitative evaluation + simulation of other sonar devices
Regarding experimental evaluation, only three of the analyzed ray-based works assessed the performance of their works, although restricted only to computational time, showing high rendering frame rates. The similarity analysis with real images demonstrated a raise with respect to our previous work for complex scenes, and comparable results for simpler scenes. In fact, visual analysis of insonification from substantial detailed scenes illustrated that the proposed simulator may help in developing and validating sonar-based intelligent techniques, such as navigation, obstacle avoidance and target tracking. Finally, our simulator was able to reproduce characteristics of sonar device operation, although more proper evaluation needs and depends on the acquisition of more real sonar data for comparison.

% future works asdas
For future works, our guess is that the use of a ray-geometry intersection algorithm along with spatial data structures, such as K-D trees and Octrees, might optimize intersection tests, mainly for dynamic and complex scenarios. Also, since the sonar simulator is open-source, next step will focus to support other robotics platforms as ROS framework\footnote{If the paper is accepted, code of the proposed simulator will become available for all the research community.}.

\FloatBarrier

\section*{References}

\bibliography{elsarticle-template}

\end{document}